\documentclass[preprint,nofootinbib,onecolumn,11pt]{revtex4}
\usepackage[tbtags]{amsmath}
\usepackage{epsfig}
\usepackage{amssymb}
\usepackage{color}
\usepackage{multirow}
\usepackage{epstopdf}
\usepackage{graphicx}
\usepackage{hyperref}
\usepackage{booktabs}

\begin{document}

\title{Discovery Potentials of Doubly Charmed Baryons}


\author{Fu-Sheng Yu$^{1,2}$}\email{yufsh@lzu.edu.cn}
\author{Hua-Yu Jiang$^{1,2}$} %
\author{Run-Hui Li$^{3}$}%
\author{Cai-Dian L\"u$^{4,5}$}\email{lucd@ihep.ac.cn}%
\author{Wei Wang$^{6}$}\email{wei.wang@sjtu.edu.cn}%
\author{Zhen-Xing Zhao$^{6}$}

\affiliation{$^{1}$ School of Nuclear Science and Technology, Lanzhou University, Lanzhou 730000, China}
\affiliation{$^{2}$ Research Center for Hadron and CSR Physics, Lanzhou University and Institute of Modern Physics of CAS, Lanzhou 730000, China}
\affiliation{$^{3}$ School of Physical Science and Technology, Inner Mongolia University, Hohhot 010021, China}
\affiliation{$^{4}$ Institute of High Energy Physics, YuQuanLu 19B, Beijing 100049, China}
\affiliation{$^{5}$ School of Physics, University of Chinese Academy of Sciences, YuQuanLu 19A, Beijing 100049, China}
\affiliation{$^{6}$ INPAC, Shanghai Key Laboratory for Particle Physics and Cosmology, School of Physics and Astronomy, Shanghai Jiao-Tong University, Shanghai 200240, China}

\begin{abstract}

The existence of doubly heavy flavor baryons has not been well established experimentally so far. In this Letter we systematically investigate the weak decays of the doubly charmed baryons, $\Xi_{cc}^{++}$ and $\Xi_{cc}^{+}$, which should be helpful for experimental searches for these particles. The long-distance contributions are first studied in the doubly heavy baryon decays, and found to be significantly enhanced. Comparing all the processes, $\Xi_{cc}^{++}\to \Lambda_c^+K^-\pi^+\pi^+$ and $\Xi_c^+\pi^+$ are the most favorable decay modes for experiments to search for doubly heavy baryons.

\end{abstract}

\maketitle

\section{Introduction}

Plenty of hadrons including quite a few exotic candidates  have been discovered in experiments during the past few decades. Doubly and triply heavy flavor baryons, however, with two or three heavy ($b$ or $c$) quarks, are so far still absent in hadron spectroscopy \cite{Klempt:2009pi,Crede:2013sze,Cheng:2015iom,Chen:2016spr}. Searches for the doubly and triply heavy baryons will play a key role in completing hadron spectroscopy and shedding light on perturbative and non-perturbative QCD dynamics.

The only evidence was reported by the SELEX experiment for $\Xi_{cc}^+$ via the process $\Xi_{cc}^{+}\to \Lambda_{c}^{+}K^{-}\pi^{+}$ in 2002 \cite{Mattson:2002vu}, followed by $\Xi_{cc}^{+}\to pD^{+}K^{-}$ \cite{Ocherashvili:2004hi}. However, this has not  been confirmed by any other experiments so far. The FOCUS experiment  reported no signal right after SELEX's measurements \cite{Ratti:2003ez}. The BaBar \cite{Aubert:2006qw} and Belle \cite{Chistov:2006zj,Kato:2013ynr} experiments searched for $\Xi_{cc}^{+(+)}$ with the final states of  $\Xi_{c}^{0}\pi^{+}(\pi^{+})$ and $\Lambda_{c}^{+}K^{-}\pi^{+}(\pi^{+})$, and did not find any evidence. The LHCb experiment performed a search using the 0.65\,fb$^{-1}$ data sample in the discovery channel used by SELEX, $\Xi_{cc}^{+}\to \Lambda_{c}^{+}K^{-}\pi^{+}$, but no significant signal was observed \cite{Aaij:2013voa}.  Besides, the mass measured by SELEX, $m_{\Xi_{cc}^{+}}=3518.9\pm0.9$\,MeV, is much lower than most theoretical predictions, for instance $m_{\Xi_{cc}}$= 3.55-3.67 GeV predicted by lattice QCD \cite{Lewis:2001iz,Flynn:2003vz,Liu:2009jc,Alexandrou:2012xk,Briceno:2012wt,Alexandrou:2014sha}. These puzzles can only be solved by experimental measurements with high luminosities.

At the Large Hadron Collider (LHC), plenty of heavy quarks have been generated, and thereby abundant doubly heavy hadrons has been produced due to quark-hadron duality, such as $B_c^{\pm}$, which has been studied in great detail by LHCb. The cross sections of the hadronic production of $\Xi_{cc}$ at the LHC have been calculated  in QCD \cite{Zhang:2011hi}, and are of the same order as those of  $B_c$ \cite{Chang:2005bf}. As LHCb has a data sample larger than 3\,fb$^{-1}$ and is collecting even more data during Run 2, there is now a good opportunity to study $\Xi_{cc}$. One key issue left is to select the decay processes with the largest possibility of observing doubly charmed baryons.

In this work, we will  systematically study the processes of $\Xi_{cc}^{++}$ and $\Xi_{cc}^{+}$ decays to find those with the largest branching fractions which should be helpful for experimental searches for the doubly charmed baryons. The lowest-lying heavy particles can only decay weakly. We will analyze the color-allowed tree-operator dominated decay modes using the factorization ansatz.  For other decay channels which are suppressed in the factorization scheme, the non-factorizable  contributions might be significant and behave as long distance contributions. With a direct calculation in the rescattering mechanism, we will demonstrate that long-distance contributions  are significantly enhanced   for some decay modes with high experimental efficiencies.
At the end,  we will point out that instead of searching for the $\Xi_{cc}$ using the SELEX discovery channels $\Xi_{cc}^+\to \Lambda_c^+K^-\pi^+$ and $pD^+K^-$, one should  measure $\Xi_{cc}^{++}\to \Lambda_{c}^{+}K^{-}\pi^{+}\pi^{+}$ and $\Xi_{c}^{+}\pi^{+}$ with the highest  priority.

The branching fractions depend on the lifetimes of $\Xi_{cc}^{++}$ and $\Xi_{cc}^{+}$ which, however, are predicted to be quite different in the literature. Predictions for the lifetime of $\Xi_{cc}^+$ vary from 53\,fs and 250\,fs, while those for $\Xi_{cc}^{++}$ range from 185\,fs to 670\,fs \cite{Karliner:2014gca,Kiselev:2001fw,Chang:2007xa,Onishchenko:2000yp} except for $\tau_{\Xi_{cc}^{++}}$=1550\,fs in Ref. \cite{Guberina:1999mx} which is too large compared to the lifetimes of singly charmed baryons. Despite the large ambiguity in the absolute lifetimes, it is expected that
$\tau(\Xi_{cc}^{++})\gg\tau(\Xi_{cc}^+)$ for their relative lifetimes, due to the effect of the destructive Pauli interference in the former. The ratio between their lifetimes is then
\begin{align}\label{eq:Rtau}
\mathcal{R}_\tau\equiv {\tau_{\Xi_{cc}^+}\over\tau_{\Xi_{cc}^{++}}} =0.25\sim0.37,
\end{align}
with small uncertainty in all these calculations \cite{Karliner:2014gca,Kiselev:2001fw,Chang:2007xa,Onishchenko:2000yp}. The branching fractions  of $\Xi_{cc}^{++}$ decays should be relatively larger  due to its longer lifetime, compared to those of $\Xi_{cc}^+$. Besides, particles with longer lifetimes can be better identified with high efficiency at the detectors. Thus, we recommend experimentalists to search for $\Xi_{cc}^{++}$ before $\Xi_{cc}^{+}$.


\section{Form factors}

In the study of the exclusive modes of heavy hadron decays, the transition form factors are required in the calculations. The hadronic matrix elements of $\Xi_{cc}$ decaying into the anti-triplet and sextet singly charmed baryons, i.e. $\mathcal{B}_c=\Xi_c$, $\Xi'_c$, $\Lambda_c$ and $\Sigma_c$, are expressed in terms of the form factors as
\begin{align}\label{eq:HMEb}
&\langle \mathcal{B}_c(p_f)|J^{\rm w}_\mu|\Xi_{cc}(p_i)\rangle
\nonumber\\
&=\bar u_f(p_f)\left[\gamma_\mu f_1(q^2)+{i\sigma_{\mu\nu}q^\nu\over m_i}f_2(q^2)+{q_\mu\over m_i}f_3(q^2)\right]u_i(p_i)
\nonumber\\
&-\bar u_f(p_f)\left[\gamma_\mu g_1(q^2)+{i\sigma_{\mu\nu}q^\nu\over m_i}g_2(q^2)+{q_\mu\over m_i}g_3(q^2)\right]\gamma_5u_i(p_i),\nonumber\\
\end{align}
where the initial and final baryons are all ${1\over2}^+$ states, $J^{\rm w}_\mu$ is the weak current in the relevant decays and $q=p_i-p_f$. In this work, the form factors are calculated in the light-front quark model (LFQM). The LFQM is a relativistic quark model under the light front approach, and has been successfully used to study the form factors of heavy meson and heavy baryon decays \cite{Ke:2007tg,Ke:2012wa,Cheng:1996if}.
We adopt the diquark picture for the two spectator quarks~\cite{Ke:2012wa,Ke:2007tg}. The diquark state with a charm quark and a light quark can be either a scalar ($J^P=0^{+}$), or an axial-vector state $(1^{+})$. Considering the wave functions of the relevant baryons, the hadronic matrix elements of $\Xi_{cc}$ decaying into the anti-triplet and sextet singly charmed baryons are linear combinations of the transitions with the scalar and the axial-vector diquarks,
\begin{align}
\begin{split}
\langle \mathcal{B}_c({\bf\overline 3}) | J^{\rm w}_\mu| \Xi_{cc} \rangle & = {\sqrt6\over4} \langle J^{\rm w}_\mu \rangle_{0^+} + {\sqrt6\over4} \langle J^{\rm w}_\mu \rangle_{1^+},
\\
\langle \mathcal{B}_c({\bf 6})  | J^{\rm w}_\mu | \Xi_{cc} \rangle  &= -{3\sqrt2\over4} \langle J^{\rm w}_\mu \rangle_{0^+} + {\sqrt2\over4} \langle J^{\rm w}_\mu \rangle_{1^+}.
 \end{split}
\end{align}
In the case with two identical quarks in the final state, for instance $\Sigma_{c}^{0}$, an overall factor of $\sqrt2$ has to be considered. The details of the calculations of the form factors in the LFQM can be seen in Ref. \cite{1707.02834}. The results are given in Table \ref{tab:formfactor}, with the $q^2$ dependence of
\begin{eqnarray}
F(q^2)={F(0)/( 1+\alpha{q^2\over m_{\rm fit}^2}+\delta {q^4\over m_{\rm fit}^4}}),
\end{eqnarray}
where $\alpha=+1$ for $g_{2}(q^{2})$ with $1^{+}$ diquarks as shown with stars in Table \ref{tab:formfactor}, and $\alpha=-1$ for all the other form factors. Under the flavor $SU(3)$ symmetry, the form factors are related to each other between $\Xi_{cc}^{++}$ and $\Xi_{cc}^{+}$ decays, and between $c\to s$ and $c\to d$  transitions as seen in Table \ref{tab:formfactor}. The uncertainties of the form factors can then be mostly cancelled in the relative branching fractions between decay channels.
%
%
%
\begin{table}[!]
\begin{center}
\caption{\label{tab:formfactor} The transition form factors of $\Xi_{cc}$  decaying into singly charmed baryons. The form factors with the scalar and the axial-vector diquarks are shown with $0^+$ and $1^+$ respectively. }
\scriptsize
\begin{tabular*}{88mm}{@{\extracolsep{\fill}}c|cccc|cccc}
\hline &  \multicolumn{4}{c|}{$\Xi_{cc}\to\Xi_c/\Xi'_c(0^+)$} & \multicolumn{4}{c}{$\Xi_{cc}\to\Xi_c/\Xi'_c(1^+)$}
 \\
  &$f_1$&$g_1$ &$f_2$ &$g_2$&$f_1$ &$g_1$  & $f_2$ & $g_2^*$
 \\
 \hline
 $F(0)$ &0.75 &0.62 &$-$0.78 & $-$0.08 & 0.74 & $-$0.20 & 0.80 & $-$0.02
  \\
 $m_{\rm fit}$ &  1.84 &  2.16 &  1.67 & 1.29 & 1.58 & 2.10 & 1.62 & 1.62
  \\
 $\delta$  & 0.25 & 0.35 & 0.30 & 0.52 & 0.36 & 0.21 & 0.31 & 1.37
 \\ \hline
 &  \multicolumn{4}{c|}{$\Xi_{cc}\to\Lambda_c/\Sigma_c(0^+)$} & \multicolumn{4}{c}{$\Xi_{cc}\to\Lambda_c/\Sigma_c(1^+)$}
 \\ 
  &$f_1$&$g_1$ &$f_2$ &$g_2$&$f_1$ &$g_1$  & $f_2$ & $g_2^*$
 \\
 \hline
 $F(0)$ &0.65 &0.53 & $-$0.74 & $-$0.05 & 0.64 & $-$0.17 & 0.73 & $-$0.03
  \\
 $m_{\rm fit}$ &  1.72 &  2.03 &  1.56 & 1.12 & 1.49 & 1.99 & 1.53 & 2.03
  \\
 $\delta$  & 0.27 & 0.38 & 0.32 & 1.10 & 0.37 & 0.23 & 0.32 & 2.62
 \\
\hline
\end{tabular*}
\end{center}
\end{table}

\section{Short-distance contribution dominated  processes}

With the form factors obtained above, we proceed to study the non-leptonic  decays of  $\Xi_{cc}$. The short-distance contributions in the external and internal $W$-emission amplitudes of two-body non-leptonic modes are calculated in the factorization approach manifested in the heavy quark limit. The amplitudes of $\Xi_{cc}$ decaying into a singly charmed baryon and a light meson $(M)$ can be expressed by the product of hadronic matrix elements
\begin{align}\label{eq:ampSD}
\begin{split}
\mathcal{A}(\Xi_{cc}&\to \mathcal{B}_{c}M)_{\rm SD}
=\lambda\langle M(q)|J^{\mu}|0\rangle \langle\mathcal{B}_{c}(p_{f})|J^{\rm w}_{\mu}|\Xi_{cc}(p_{i})\rangle,
\end{split}
\end{align}
where $\lambda={G_F\over\sqrt2}V_{CKM} a_{1,2}(\mu)$, $V_{CKM}$ denotes the product of the corresponding Cabibbo-Koboyashi-Maskawa matrix elements, and $a_1(\mu)=C_1(\mu)+C_2(\mu)/3$ for the external $W$-emission amplitudes  and $a_2(\mu)=C_2(\mu)+C_1(\mu)/3$ for the internal $W$-emission ones, with $C_1(\mu)=1.21$ and $C_2(\mu)=-0.42$ at the scale of  $\mu=m_c$ \cite{Li:2012cfa}. In this work $M$ denotes a pseudoscalar meson $(P)$ or a vector meson $(V)$, with the hadronic matrix elements of decay constants as
\begin{eqnarray}
\langle P (q) | J^\mu |0\rangle  = i f_{P}q^{\mu}, \;\; \langle V (q) | J^\mu |0\rangle  = f_{V}m_{V}\epsilon^{\mu*}
\end{eqnarray}
In this work, we show the results of a few gold channels with the highest probability of being observed  in experiments.
More discussions on various  processes can be found  in Refs.~\cite{Li:2017ndo,1707.02834,Wang:2017azm,Shi:2017dto}. According to Eq. (\ref{eq:ampSD}), the relative branching fractions of the other processes  compared to that of $\Xi_{cc}^{++}\to \Xi_{c}^{+}\pi^{+}$ are given as
\begin{align}\label{eq:BrSD}
&{\mathcal{B}(\Xi_{cc}^{+}\to\Xi_{c}^{0}\pi^{+}) / \mathcal{B}(\Xi_{cc}^{++}\to\Xi_{c}^{+}\pi^{+}) }=\mathcal{R}_{\tau}=0.25\sim0.37,
\nonumber\\
&{\mathcal{B}(\Xi_{cc}^{++}\to\Lambda_{c}^{+}\pi^{+}) / \mathcal{B}(\Xi_{cc}^{++}\to\Xi_{c}^{+}\pi^{+})}=0.056,
\\
&\mathcal{B}(\Xi_{cc}^{++}\to\Xi_{c}^{+}\ell^{+}\nu)/\mathcal{B}(\Xi_{cc}^{++}\to\Xi_{c}^{+}\pi^{+})=0.71,\nonumber
\end{align}
The above relations are basically unambiguous, since the uncertainties from the transition form factors are mostly cancelled under the flavor $SU(3)$ symmetry.  It is obvious that the branching fraction of $\Xi_{cc}^{++}\to \Xi_c^+\pi^+$ is the largest, compared to that of $\Xi_{cc}^{++}\to\Lambda_{c}^{+}\pi^{+}$ which is a Cabibbo-suppressed mode, and that of $\Xi_{cc}^{+}\to \Xi_c^0\pi^+$ due to the expected smaller lifetime of $\Xi_{cc}^+$. The semi-leptonic mode of $\Xi_{cc}^{++}\to\Xi_{c}^{+}\ell^{+}\nu$ suffers a low efficiency of detection for the missing energies of neutrinos. Similarly, some other processes with possible larger branching fractions may lose lots of events at hadron colliders, due to neutral particles such as in $\Xi_c^+\rho^+(\to \pi^+\pi^0)$ and $\Xi_c^{\prime+}(\to\Xi_c^+\gamma)\pi^+$. The final state of $\Xi_c^+a_1^+(\to\pi^+\pi^+\pi^-)$ has two more tracks, reducing the efficiency of detection. Besides, the longer lifetimes of $\Xi_{cc}^{++}$ and $\Xi_c^+$ compared to those of $\Xi_{cc}^{+}$ and $\Xi_c^0$, respectively, benefit higher efficiencies of the identification of the particles in experiments. Thus the $\Xi_{cc}^{++}\to \Xi_c^+\pi^+$ process is the best of the external $W$-emission processes to search for the doubly charmed baryons.

The absolute branching fraction of $\Xi_{cc}^{++}\to \Xi_{c}^{+}\pi^{+}$ is calculated to be
\begin{align}
\mathcal{B}(\Xi_{cc}^{++}\to \Xi_{c}^{+}\pi^{+}) &=
 \left({\tau_{\Xi_{cc}^{++}}\over 300\,\text{fs}}\right)\times7.2\%.
\end{align}
This result is given by comparing the lifetime of $\Xi_{cc}^{++}$ with 300\,fs, which is in the range of the predictions. Even if considering the uncertainties of the transition form factors and the lifetime, the branching fraction of this process is of the order of percent, which is large enough for measurements.

To measure $\Xi_{cc}^{++}\to \Xi_c^+\pi^+$, $\Xi_c^+$ can be reconstructed  using the mode $\Xi_c^+\to p K^- \pi^+$ at hadron colliders with all the charged particles in the final state. The absolute branching fraction of this process has never been directly measured, but the relative branching ratio was measured as
$\mathcal{B}(\Xi_c^+\to p\overline K^{*0})/\mathcal{B}( \Xi_c^+\to p K^- \pi^+)=0.54\pm0.10$ \cite{Link:2001rn}. Besides, the relation $\mathcal{A}(\Xi_c^+\to p\overline K^{*0})=\mathcal{A}(\Lambda_c^+\to \Sigma^+ K^{*0})$ holds under $U$-spin symmetry.
With the measurement of $\mathcal{B}(\Lambda_c^+\to \Sigma^+ K^{*0})=(0.36\pm0.10)\%$ \cite{Link:2002zx}, the branching fraction is
\begin{align}\label{eq:BrXic}
\mathcal{B}( \Xi_c^+\to p K^- \pi^+)=(2.2\pm0.8)\%.
\end{align}
The relatively larger branching fraction of this Cabibbo-suppressed mode is induced by the larger phase space of $\Xi_c^+\to p \overline K^{*0}$ and the longer lifetime of $\Xi_c^+$.  The main uncertainty in Eq. \eqref{eq:BrXic} arises from the branching fraction of $\Lambda_c^+\to \Sigma^+K^{*0}$ and the ratio between $\Xi_c^+\to p\overline K^{*0}$ and $\Xi_c^+\to p K^- \pi^+$, which may be measured by BESIII, Belle II and LHCb with higher precision. Considering the relatively large value of $\mathcal{B}( \Xi_c^+\to p K^- \pi^+)$ within the $1\sigma$ range, we suggest to measure the process of $\Xi_{cc}^{++}\to\Xi_{c}^{+}\pi^{+}$ with $\Xi_{c}^+$ reconstructed by the final state $p K^- \pi^+$.

\section{Long-distance contribution dominated processes}

In the factorization approach, only the factorizable contributions are taken into account. For the color-allowed  tree-operator dominated channels, the non-factorizable contributions are expected to be small. For the color-suppressed processes with a tiny Wilson coefficient $a_2$, the decay widths are likely to be underestimated in the factorization framework.
 For instance, the branching fractions of the internal $W$-emission decays of $\Xi_{cc}^{++}\to\Sigma_{c}^{++}(2455)\overline K^{*0}$ and $\Xi_{cc}^{+}\to \Lambda_{c}^{+}\overline K^{*0}$ are predicted to be of the order of $10^{-5}$, due to   $a_{2}(\mu)\approx-0.02$.

 However, the long-distance contributions are usually significantly enhanced in charmed meson decays, which can be described well by the rescattering mechanism of the final-state-interaction effects \cite{Ablikim:2002ep,Li:2002pj,Fajfer:2003ag,Li:1997vu}. The rescattering mechanism in the heavy-flavor-baryon decays was only considered in Ref. \cite{Chen:2002jr} to study the Cabibbo-suppressed decays of $\Lambda_c^+\to p\pi^0$ and $n\pi^+$, whose results have not been directly manifested so far, but are consistent with the upper limit recently measured by BESIII \cite{Ablikim:2017ors}. In doubly heavy flavor baryon decays, the long-distance contributions have never been considered. In this work we first calculate the rescattering effects in two-body non-leptonic $\Xi_{cc}$ decays for the internal $W$-emission and $W$-exchange amplitudes, and then find some other processes with large branching fractions.

\begin{figure}
\begin{center}
\includegraphics[width=6cm]{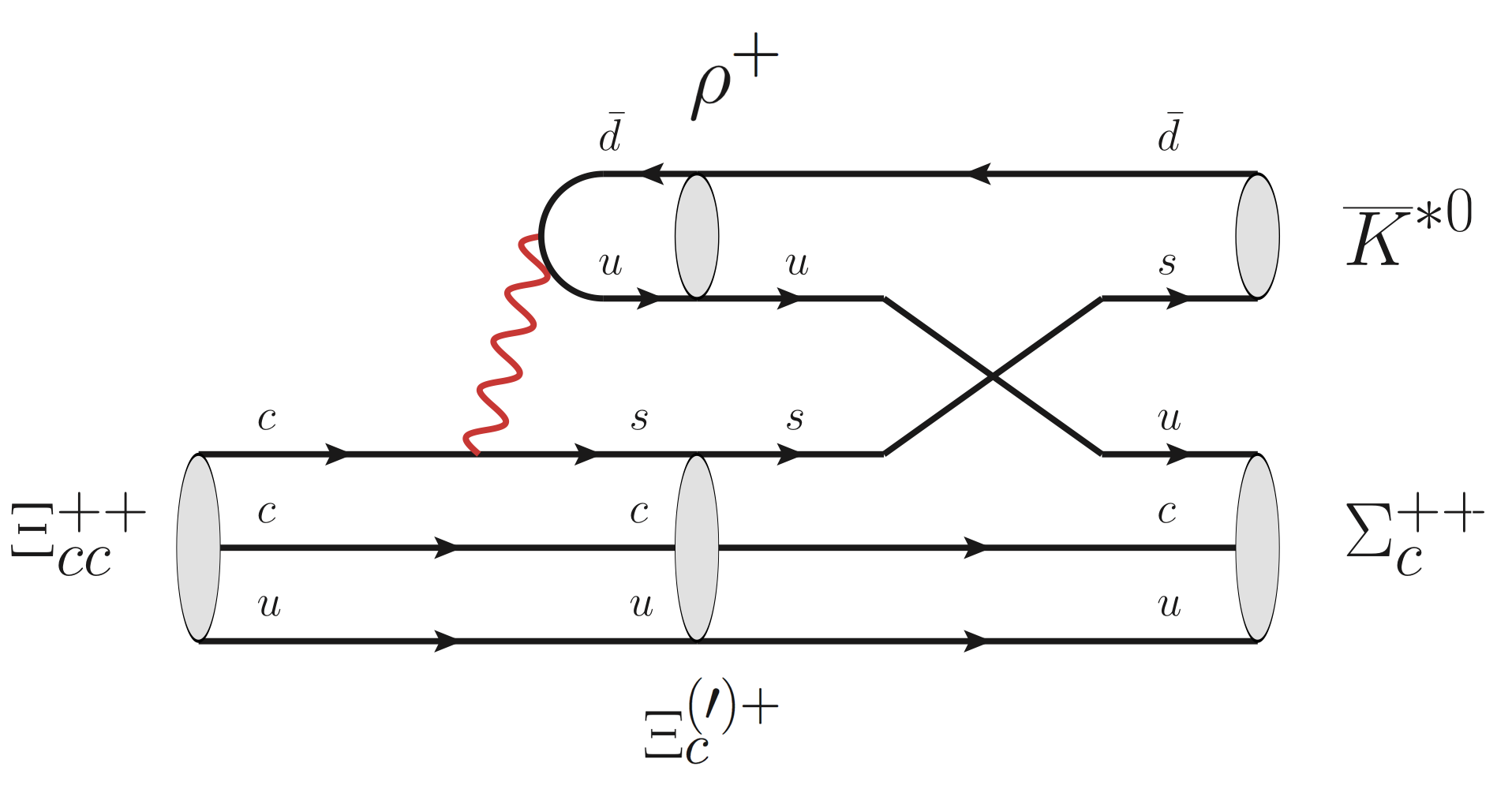}
\includegraphics[width=5cm]{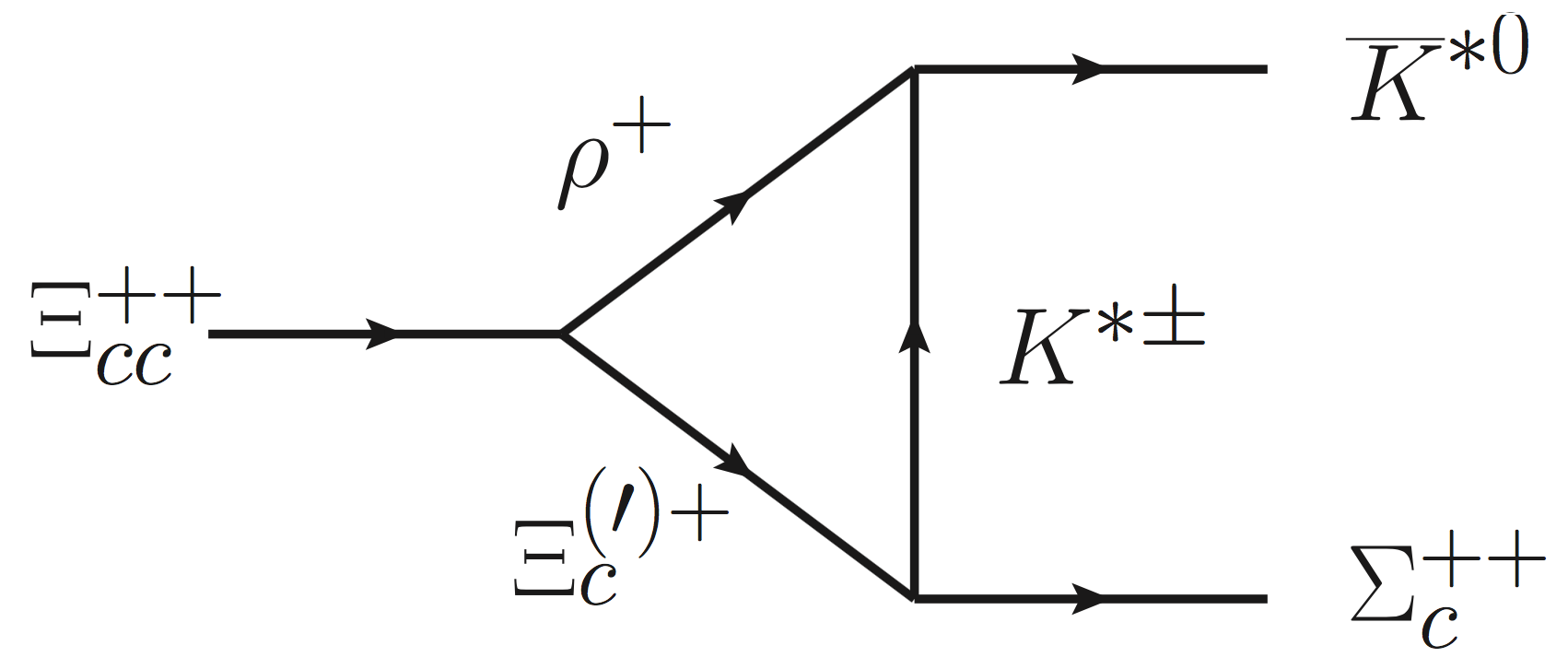}
\caption{\label{fig:triangle} The $t$-channel rescattering diagram of $\Xi_{cc}^{++}\to \Xi_{c}^{(\prime)+}\rho^{+}\to\Sigma_c^{++}\overline K^{*0}$. The rescattering is induced by the quark exchanges, shown in the top figure. The bottom figure is the diagram at the hadron level.}
\end{center}
\end{figure}

\begin{figure}
\begin{center}
\includegraphics[width=7.5cm]{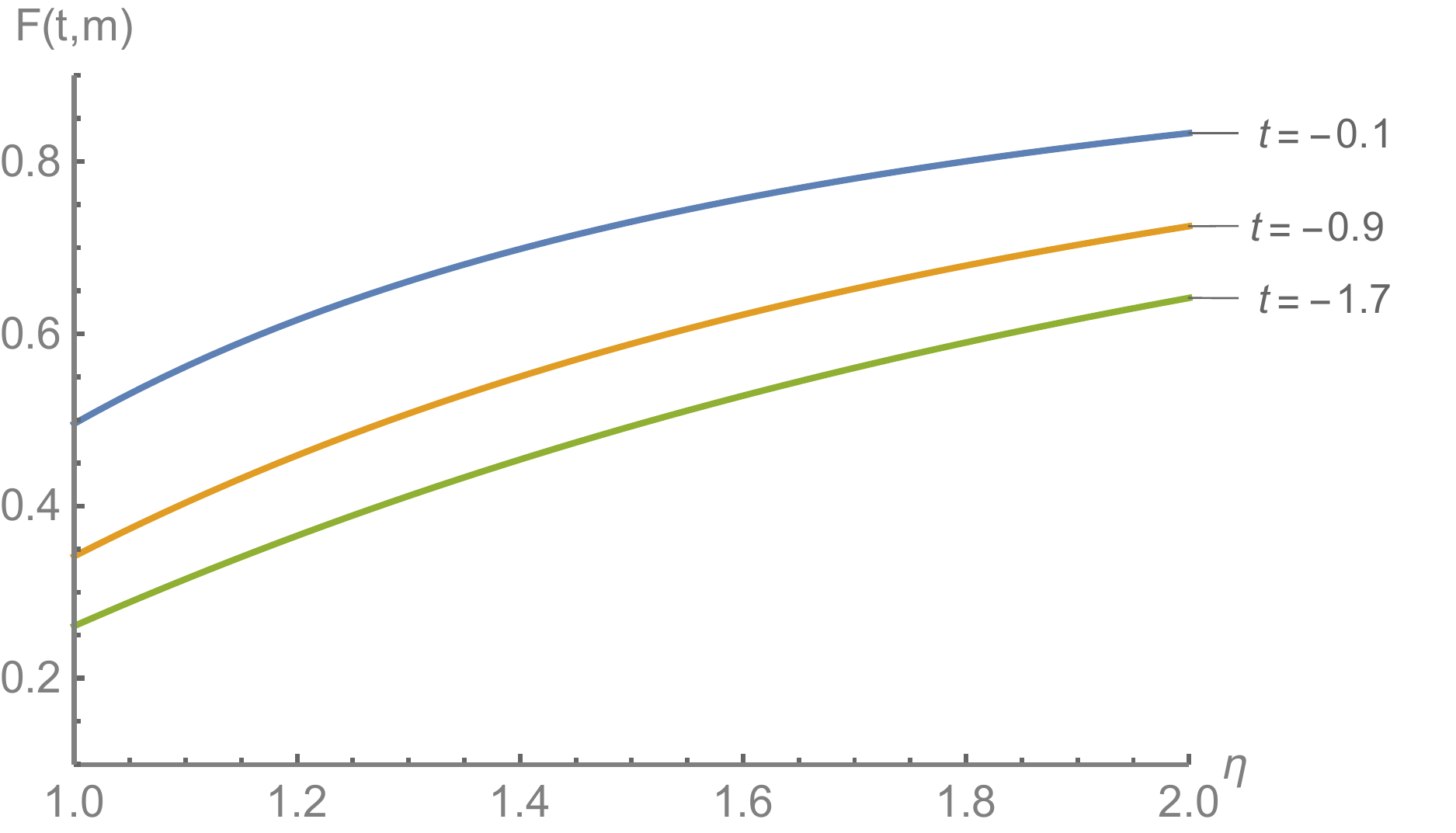}\\
\caption{\label{fig:formfactor}The form factor $F(t,m)$ as a function of $\eta$.}
\end{center}
\end{figure}

The absorptive part of the amplitudes is obtained by the optical theorem \cite{Cheng:2004ru}, summing over all possible amplitudes of $\Xi_{cc}(p_{i})$ decaying into the states $\{p_k\}$, followed by the rescattering of $\{p_k\}$ into the final state $\mathcal{B}_{c}(p_{f})M(q)$,
\begin{align}\label{eq:abs}\begin{split}\text{Abs}\mathcal{M}(p_i\to p_f q)={1\over2}\sum_j\left(\prod_{k=1}^j\int{d^3p_k\over(2\pi)^32E_k}\right)(2\pi)^4\\\times \delta^4(p_f+q-\sum_{k=1}^jp_k)\mathcal{M}(p\to \{p_k\})T^*(p_fq\to \{p_k\}).\end{split}\end{align}
One typical rescattering diagram is given in Fig. \ref{fig:triangle}, taking as example the $t$-channel triangle diagram of $\Xi_{cc}^{++}\to \Xi_{c}^{(\prime)+}\rho^{+}\to\Sigma_c^{++}\overline K^{*0}$ via quark exchange.
%
%
The rescattering amplitudes are calculated using the effective Lagrangian \cite{Yan:1992gz,Casalbuoni:1996pg,Meissner:1987ge,Li:2012bt}. The hadronic strong coupling constants are related to each other under the flavor $SU(3)$ symmetry, and the chiral and heavy quark symmetries \cite{Yan:1992gz,Casalbuoni:1996pg,Meissner:1987ge}, with the values taken from Refs. \cite{Aliev:2010yx,Aliev:2010nh,Khodjamirian:2011jp,Azizi:2014bua,Yu:2016pyo,Azizi:2015tya,Cheng:2004ru,Ballon-Bayona:2017bwk}. The effective Lagrangian and the strong coupling constants are given in the Appendix. Most of the uncertainties will then be cancelled in the relative branching ratios. The results of the rescattering amplitudes depend on the form factor $F(t,m)$ which describes the off-shell effect of the exchanged particle. It is parametrized as
$F(t,m)={(\Lambda^2-m^2) / (\Lambda^2-t)}$ \cite{Cheng:2004ru},
with the cutoff $\Lambda=m+\eta\Lambda_{\rm QCD}$, $m$ and $t$ being the mass and the momentum squared of the exchanged particle, respectively, and $\Lambda_{\rm QCD}$ taken as 330\,MeV. The free parameter $\eta$ cannot be calculated from first principles. In this work, we take $\eta$ varying in the range from 1.0 to 2.0, as found in Ref. \cite{Cheng:2004ru}. The dependence of $F(t,m)$ on $\eta$ is plotted in Fig.\ref{fig:formfactor}.

\begin{figure}
\begin{center}
\includegraphics[width=7.2cm]{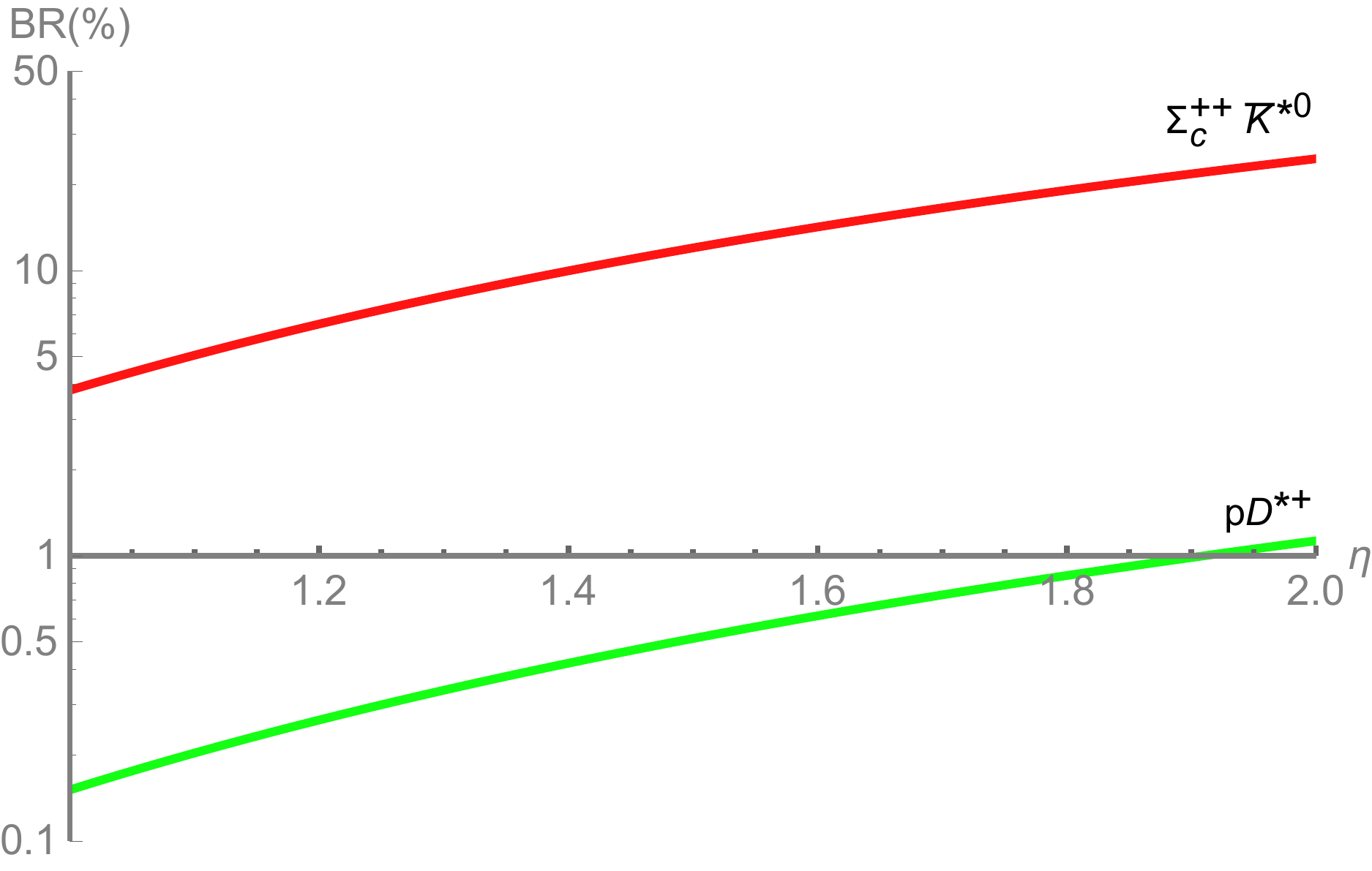}
\includegraphics[width=7.5cm]{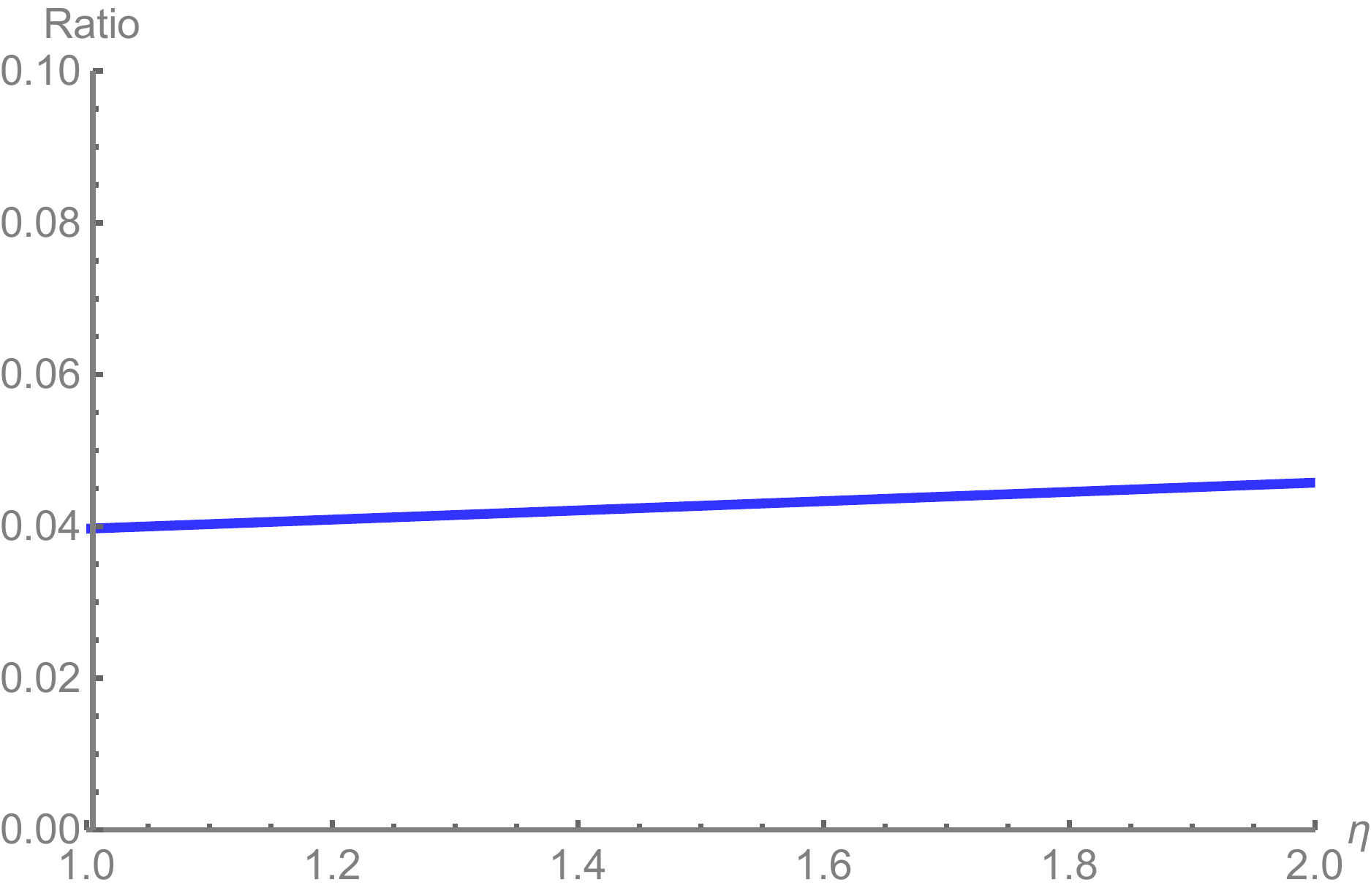}
\caption{\label{fig:BRatio}The absolute branching fractions of $\Xi_{cc}^{++}\to\Sigma_c^{++}\overline K^{\ast0}$ and $pD^{\ast+}$ (top) and the ratio of  ${\mathcal{B}(\Xi_{cc}^{++}\to pD^{\ast+})}/{\mathcal{B}(\Xi_{cc}^{++}\to\Sigma_c^{++}\overline K^{\ast0})}$ (bottom) as functions of $\eta$.}
\end{center}
\end{figure}

The relative branching fractions of some processes dominated by the long-distance contributions compared to that of $\Xi_{cc}^{++}\to\Sigma_{c}^{++}(2455)\overline K^{*0}$ are shown in Table \ref{tab:BrLD}.  The results of the relative branching fractions are less ambiguous in theory, since the uncertainties from the effective hadronic strong coupling constants and from the transition form factors are mainly cancelled due to the flavor $SU(3)$ symmetry and the chiral and heavy quark symmetries as discussed before. The absolute branching fractions depend heavily on the values of the parameter $\eta$, as seen in the top plot of Fig. \ref{fig:BRatio}, taking as examples $\mathcal{B}(\Xi_{cc}^{++}\to\Sigma_c^{++}(2455)\overline K^{\ast0})$ and $\mathcal{B}(\Xi_{cc}^{++}\to pD^{\ast+})$ as functions of $\eta$.  In the bottom of Fig.  \ref{fig:BRatio}, we plot the ratio ${\mathcal{B}(\Xi_{cc}^{++}\to pD^{\ast+})}/{\mathcal{B}(\Xi_{cc}^{++}\to\Sigma_c^{++}\overline K^{\ast0})}$ as a function of $\eta$. The ratio of branching fractions is insensitive to $\eta$. Therefore, the theoretical uncertainties are under control for the relative branching fractions. From Table \ref{tab:BrLD}, it is obvious that $\Xi_{cc}^{++}\to \Sigma_{c}^{++}(2455)\overline K^{*0}$ has the largest branching fraction, which is useful for experimental measurements.
\begin{table}
\begin{center}
\caption{\label{tab:BrLD} Branching fractions of $\Xi_{cc}^{++}$ and $\Xi_{cc}^+$ decays with the long-distance contributions, relative to that of $\Xi_{cc}^{++}\to \Sigma_{c}^{++}(2455)\overline K^{*0}$.}
\begin{tabular*}{84mm}{@{\extracolsep{\fill}}ccc}
\hline
Baryons & \quad\quad\quad\quad Modes \quad\quad\quad\quad & $\mathcal{B}_{\rm LD}$
\\
\hline
$\Xi_{cc}^{++}(ccu)$ &$\Sigma_{c}^{++}(2455)\overline K^{*0}$ &  defined as 1
\\
&$pD^{*+}$  & $0.04$
\\
&$pD^+$  & $0.0008$
\\
\hline
$\Xi_{cc}^+(ccd)$ &$\Lambda_c^+\overline K^{*0}$ &  $(\mathcal{R}_{\tau}/0.3) \times 0.22$
\\
&$\Sigma_c^{++}(2455)K^-$ &  $(\mathcal{R}_{\tau}/0.3) \times 0.01$
\\
&$\Xi_c^+\rho^0$ &  $(\mathcal{R}_{\tau}/0.3) \times 0.04$
\\
& $\Lambda D^+$ & $(\mathcal{R}_{\tau}/0.3) \times 0.004$
\\
&$pD^0$ &  $(\mathcal{R}_{\tau}/0.3) \times 0.001$
\\
\hline
\end{tabular*}
\end{center}
\end{table}
%
In the process $\Xi_{cc}^{++}\to\Sigma_{c}^{++}(2455)\overline K^{*0}$, the dominant rescattering amplitude is $\Xi_{cc}^{++}\to \Xi_c^{(\prime)+}\rho^+\to\Sigma_c^{++} \overline K^{*0}$ with exchange of $K^{*\pm}$, depicted in Fig. \ref{fig:triangle}. Considering some other triangle diagrams including the $t$-channel rescattering by  $\Xi_c^{(\prime)+}\pi^+$ with exchange of $K^\pm$, and $u$-channels rescattering by $\Xi_c^{(\prime)+}\pi^+$ and $\Xi_c^{(\prime)+}\rho^+$ with exchanging $\Lambda_c^+$ or $\Sigma_c^+$,
the absolute branching fraction is
\begin{align}\label{eq:Brbenchmark}
\mathcal{B}(\Xi_{cc}^{++}\to\Sigma_{c}^{++}(2455)\overline K^{*0})=\left({\tau_{\Xi_{cc}^{++}}\over 300\,\text{fs}}\right)\times(3.8\sim24.6)\%,
\end{align}
where the range corresponds to the value of $\eta$ varying between 1.0 and 2.0. Compared to the short-distance results of $\Xi_{cc}^{++}\to\Sigma_{c}^{++}(2455)\overline K^{*0}$ as $\mathcal{O}(10^{-5})$, the long-distance contributions in the doubly charmed baryon decays are significantly enhanced. The branching fractions of $\Xi_{cc}^{++}\to \Xi_c^+\rho^+$ and $\Xi_c^{\prime+}\rho^+$ are $12.6\%$ and $17.4\%$ respectively, which are large enough to lead to a result for $\Xi_{cc}^{++}\to\Sigma_{c}^{++}(2455)\overline K^{*0}$ of the order of percent. In the charmed meson decays, the large-$N_c$ approach gives a good description of the internal $W$-emission contributions, which amounts to $|a_{2}^{\rm eff}(\mu_c)|\approx C_2(\mu_c)\sim-0.5$ \cite{Cheng:2010ry}. With this value, the branching fraction of this process is then $4.6\%$, which is in the range of Eq. \eqref{eq:Brbenchmark}. So the results considering the rescattering mechanism for long-distance contributions are trustworthy.

$\Xi_{cc}^{++}\to \Sigma_c^{++}(2455)\overline K^{*0}$ is actually a four-body process with the strong decays of $\Sigma_c^{++}\to \Lambda_c^+\pi^+$ and $\overline K^{*0}\to K^-\pi^+$. In the charmed meson decays, the resonant contributions almost saturate the decay widths \cite{Cheng:2010rv}. The final-state particles are not very energetic in charm decays, and hence easily located in the momentum range of resonances. Thus the resonant contributions can indicate the key physics. For the four-body process $\Xi_{cc}^{++}\to \Lambda_{c}^{+}K^{-}\pi^{+}\pi^{+}$, there are many low-lying-resonance contributions, such as $\Sigma_c^{++}(2455)$ and $\Sigma_c^{++}(2520)$ for $\Lambda_c^+\pi^+$, and $\overline K^{*0}$ and $(K\pi)_{\rm S\text{-}wave}$ for $K^-\pi^+$. Recalling that $\mathcal{B}(\Xi_{cc}^{++}\to \Sigma_{c}^{++}(2455)\overline K^{*0})$ is as large as shown in Eq.(\ref{eq:Brbenchmark}), the branching fraction of $\Xi_{cc}^{++}\to \Lambda_{c}^{+}K^{-}\pi^{+}\pi^{+}$ will be expected to be a few percent, or even reach $\mathcal{O}(10\%)$. With $\Lambda_c^+$ reconstructed by $pK^-\pi^+$, the process $\Xi_{cc}^{++}\to \Lambda_{c}^{+}K^{-}\pi^{+}\pi^{+}$ can be used to search for the doubly charmed baryon.

Apart from the above four-body decay mode, the branching fraction of $\Xi_{cc}^{+}\to\Lambda_c^+K^-\pi^+$ should be considerable, contributed by $\Lambda_c^+\overline K^{*0}$ and $\Sigma_c^{++}(2455)K^-$. This process is just the one used by the SELEX experiment to report the first would-be evidence of $\Xi_{cc}^+$ \cite{Mattson:2002vu}. However, no significant signal was observed by the LHCb experiment using this channel \cite{Aaij:2013voa}. We find the process $\Xi_{cc}^{++}\to \Lambda_{c}^{+}K^{-}\pi^{+}\pi^{+}$ is better than $\Xi_{cc}^{+}\to\Lambda_c^+K^-\pi^+$ for the searches for doubly charmed baryons, for the following reasons. For the dominant resonant contributions in these two processes, the branching fraction of $\Xi_{cc}^{++}\to \Sigma_{c}^{++}(2455)\overline K^{*0}$ is larger than that of $\Xi_{cc}^{+}\to\Lambda_c^+\overline K^{*0}$ by a factor of about five, due to the predicted value of $\mathcal{R}_\tau\sim0.3$ with a small uncertainty, seen in Eq. \eqref{eq:Rtau}. As explained before, the efficiency of identifying $\Xi_{cc}^{++}$ is larger than that of $\Xi_{cc}^{+}$ at LHCb by a factor around the lifetime ratio, in the range of their predicted lifetimes. Even though $\Xi_{cc}^{++}\to \Lambda_{c}^{+}K^{-}\pi^{+}\pi^{+}$ suffers a lower efficiency in detection by a few factors, due to one more track than $\Xi_{cc}^{+}\to\Lambda_c^+K^-\pi^+$, it can still be expected that there would be more signal yields in the former process. For the other discovery channel by the SELEX experiment, $\Xi_{cc}^+\to pD^+K^-$, there are no low-lying-resonance contributions. In the mode $\Xi_{cc}^+\to \Lambda D^+$, whose branching ratio is small, the $\Lambda$ state is below the $pK^-$ threshold, while the higher excited resonances would be more difficult to produce. Therefore, the process $\Xi_{cc}^{++}\to \Lambda_{c}^{+}K^{-}\pi^{+}\pi^{+}$ is the best of the long-distance contribution dominated processes for the searches for doubly charmed baryons.

In addition to the study of the ground states of doubly charmed baryons, the suggested processes should be useful to search for excited states below the charm-meson-charm-baryon thresholds. Such particles strongly or radiatively decay into the ground states, which would be reconstructed in experiments by the most favorable modes found in this work. Besides, the long-distance contributions, for which we found large and important $\Xi_{cc}$ decays, should also be considered in the studies on the search for the discovery channels of other heavy particles, such as bottom-charm baryons and stable open flavor tetraquarks and pentaquarks.


\section{Summary}
We have systematically studied the weak decays of $\Xi_{cc}^{++}$ and $\Xi_{cc}^{+}$ and recommend the processes $\Xi_{cc}^{++}\to \Lambda_{c}^{+}K^{-}\pi^{+}\pi^{+}$ and $\Xi_{cc}^+\pi^+$ as the most favorable decay modes for searches for doubly charmed baryons in experiments. The channels $\Xi_{cc}^{+}\to\Lambda_{c}^{+}K^{-}\pi^{+}$ and $pD^{+}K^{-}$ used by the SELEX and LHCb experiments are not as good as the above two $\Xi_{cc}^{++}$ decay processes. The short-distance contributions of the decay amplitudes are calculated under the factorization approach. The long-distance contributions are first studied in the double-charm-baryon decays, considering the rescattering mechanism. It is found that the long-distance contributions are significantly enhanced and are essential for the favorable mode $\Xi_{cc}^{++}\to \Lambda_{c}^{+}K^{-}\pi^{+}\pi^{+}$. Our suggestions are based on the analysis of the relative branching fractions between decay modes, which is less ambiguous since the theoretical uncertainties are mainly cancelled by the flavor symmetries. The absolute branching fractions of $\Xi_{cc}^{++}\to \Lambda_c^+K^-\pi^+\pi^+$ and $\Xi_c^+\pi^+$ are estimated to be a few percent, or even reach the order of $10\%$, which are large enough for experimental measurements.
\footnotetext[1]
{Note added:
very recently, the LHCb collaboration reported the discovery of $\Xi_{cc}^{++}$ with the final state $\Lambda_{c}^{+}K^{-}\pi^{+}\pi^{+}$ \cite{Aaij:2017ueg}.}

\acknowledgments{We are grateful for Ji-Bo He for enlightening discussions which initiated this project, and Hai-Yang Cheng and Xiang Liu for careful proofreadings.}

\appendix{\bf Appendix:} 
The effective Lagrangians used in the rescattering mechanism are \cite{Yan:1992gz,Casalbuoni:1996pg,Meissner:1987ge,Li:2012bt}:
\begin{align}
\begin{split}\label{eq:StrongLag}
\mathcal{L}_{eff}=&\mathcal{L}_{\pi hh}+\mathcal{L}_{\rho hh}+\mathcal{L}_{\pi\mathcal{B}\mathcal{B}}+\mathcal{L}_{\rho \mathcal{B}\mathcal{B}}+\mathcal{L}_{\rho\pi\pi}
+\mathcal{L}_{\rho\rho\rho}+\mathcal{L}_{\rho DD}+\mathcal{L}_{\pi D^{\ast}D}+\mathcal{L}_{\rho D^{\ast}D^{\ast}},\\
\mathcal{L}_{\pi hh}=&g_{\pi \mathcal{B}_{6}\mathcal{B}_{6}}Tr\lbrack\bar{\mathcal{B}}_{6}i\gamma_{5}\Pi \mathcal{B}_{6}\rbrack+g_{\pi \mathcal{B}_{\bar{3}}\mathcal{B}_{\bar{3}}}Tr\lbrack\bar{\mathcal{B}}_{\bar{3}}i\gamma_{5}\Pi \mathcal{B}_{\bar{3}}\rbrack+\{g_{\pi \mathcal{B}_{6}\mathcal{B}_{\bar{3}}}Tr\lbrack\bar{\mathcal{B}}_{6}i\gamma_{5}\Pi \mathcal{B}_{\bar{3}}\rbrack+h.c.\},\\
\mathcal{L}_{\rho hh}=&f_{1\rho \mathcal{B}_{6}\mathcal{B}_{6}}Tr[\bar{\mathcal{B}}_{6}\gamma_{\mu}V^{\mu}\mathcal{B}_{6}]+\frac{f_{2\rho \mathcal{B}_{6}\mathcal{B}_{6}}}{2m_{6}}Tr[\bar{\mathcal{B}}_{6}\sigma_{\mu\nu}\partial^{\mu}V^{\nu}\mathcal{B}_{6}]\\
&+f_{1\rho \mathcal{B}_{\bar{3}}\mathcal{B}_{\bar{3}}}Tr[\bar{\mathcal{B}}_{\bar{3}}\gamma_{\mu}V^{\mu}\mathcal{B}_{\bar{3}}]+\frac{f_{2\rho \mathcal{B}_{\bar{3}}\mathcal{B}_{\bar{3}}}}{2m_{\bar{3}}}Tr[\bar{\mathcal{B}}_{\bar{3}}\sigma_{\mu\nu}\partial^{\mu}V^{\nu}\mathcal{B}_{\bar{3}}]\\
&+\{f_{1\rho \mathcal{B}_{6}\mathcal{B}_{\bar{3}}}Tr[\bar{\mathcal{B}}_{6}\gamma_{\mu}V^{\mu}\mathcal{B}_{\bar{3}}]+\frac{f_{2\rho \mathcal{B}_{6}\mathcal{B}_{\bar{3}}}}{m_{6}+m_{\bar{3}}}Tr[\bar{\mathcal{B}}_{6}\sigma_{\mu\nu}\partial^{\mu}V^{\nu}\mathcal{B}_{\bar{3}}]+h.c.\},\\
\mathcal{L}_{\pi \mathcal{B}\mathcal{B}}=&g_{\pi \mathcal{B}\mathcal{B}}Tr[\bar{\mathcal{B}}i\gamma_{5}\Pi \mathcal{B}],\\
\mathcal{L}_{\rho \mathcal{B}\mathcal{B}}=&f_{1\rho \mathcal{B}\mathcal{B}}Tr[\bar{\mathcal{B}}\gamma_{\mu}V^{\mu}\mathcal{B}]+\frac{f_{2\rho \mathcal{B}\mathcal{B}}}{2m_{\mathcal{B}}}Tr[\bar{\mathcal{B}}\sigma_{\mu\nu}\partial^{\mu}V^{\nu}\mathcal{B}],\\
\mathcal{L}_{\rho\pi\pi}=&\frac{ig_{\rho\pi\pi}}{\sqrt{2}}Tr[V^{\mu}[\Pi,\partial_{\mu}\Pi]],\\
\mathcal{L}_{\rho\rho\rho}=&\frac{ig_{\rho\rho\rho}}{\sqrt{2}}Tr[(\partial_{\nu}V_{\mu}-\partial_{\mu}V_{\nu})V^{\mu}V^{\nu}]
=\frac{ig_{\rho\rho\rho}}{\sqrt{2}}Tr[(\partial_{\nu}V_{\mu}V^{\mu}-V^{\mu}\partial_{\nu}V_{\mu})V^{\nu}],\\
\mathcal{L}_{\rho DD}=&-ig_{\rho DD}(D_{i}\partial_{\mu}D^{j\dagger}-\partial_{\mu}D_{i}D^{j\dagger})(V^{\mu})^{i}_{j},\\
\mathcal{L}_{\pi D^{\ast}D\ }&=-g_{\pi D^{\ast}D}(D^{i}\partial^{\mu}\Pi_{ij}D_{\mu}^{\ast j\dagger}+D_{\mu}^{\ast i}\partial^{\mu}\Pi_{ij}D^{j\dagger}),\\
\mathcal{L}_{\rho D^{\ast}D^{\ast}}&=ig_{\rho D^{\ast}D^{\ast}}(D_{i}^{\ast\nu}\partial_{\mu}D_{\nu}^{\ast j\dagger}-\partial_{\mu}D_{i}^{\ast\nu}D_{\nu}^{\ast j\dagger})(V^{\mu})^{i}_{j}+4if_{\rho D^{\ast}D^{\ast}}D_{i\mu}^{\ast\dagger}(\partial^{\mu}V^{\nu}-\partial^{\nu}V^{\mu})^{i}_{j}D_{\nu}^{\ast j}.
\end{split}
\end{align}

where the corresponding $\Pi$, $V$, $\mathcal{B}$ and $\mathcal{B}_{6}$, $\mathcal{B}_{\bar{3}}$ respectively represent the matrices
\begin{align}
\begin{split}
\Pi=&
\left(
  \begin{array}{ccc}
   \frac{\pi^{0}}{\sqrt{2}}+\frac{\eta}{\sqrt{6}} & \pi^{+} & K^{+} \\
    \pi^{-} & -\frac{\pi^{0}}{\sqrt{2}}+\frac{\eta}{\sqrt{6}} & K^{0} \\
    K^{-} & \bar{K}^{0} & -\sqrt{\frac{2}{3}}\eta \\
  \end{array}
\right),
\mathcal{B}_{6}=
\left(
  \begin{array}{ccc}
   \Sigma_{c}^{++}~~ & \frac{1}{\sqrt{2}}\Sigma_{c}^{+}~~ & \frac{1}{\sqrt{2}}\Xi_{c}^{\prime+} \\
   \frac{1}{\sqrt{2}}\Sigma_{c}^{+} & \Sigma_{c}^{0} &  \frac{1}{\sqrt{2}}\Xi_{c}^{\prime0} \\
   \frac{1}{\sqrt{2}}\Xi_{c}^{\prime+} & \frac{1}{\sqrt{2}}\Xi_{c}^{\prime0} & \Omega_{c} \\
  \end{array}
\right),\\
\\
V=&
\left(
  \begin{array}{ccc}
   \frac{\rho^{0}}{\sqrt{2}}+\frac{\omega}{\sqrt{2}} & \rho^{+} & K^{\ast+} \\
    \rho^{-} & -\frac{\rho^{0}}{\sqrt{2}}+\frac{\omega}{\sqrt{2}} & K^{\ast0} \\
    K^{\ast-} & \bar{K}^{\ast0} &  \phi \\
  \end{array}
\right),
\quad \mathcal{B}_{\bar{3}}=
\left(
  \begin{array}{ccc}
   0 & ~~~~~\Lambda_{c}^{+} & ~~~~~\Xi_{c}^{+} \\
   -\Lambda_{c}^{+} & ~~~~~0 & ~~~~~\Xi_{c}^{0} \\
   -\Xi_{c}^{+} & ~~~~-\Xi_{c}^{0} & ~~~~~0 \\
  \end{array}
\right),\\
\\
\mathcal{B}=&
\left(
  \begin{array}{ccc}
   \frac{\Sigma^{0}}{\sqrt{2}}+\frac{\Lambda}{\sqrt{6}} & \Sigma^{+} & p \\
    \Sigma^{-} & -\frac{\Sigma^{0}}{\sqrt{2}}+\frac{\Lambda}{\sqrt{6}} & n \\
    \Xi^{-} & \Xi^{0} & -\frac{2}{\sqrt{6}}\Lambda \\
  \end{array}
\right).
\end{split}
\end{align}

According to the generalized form of baryons coupled with mesons in Eq.\eqref{eq:StrongLag}, we extend to the vertex $\mathcal{B}_c\mathcal{B}D$ and $\mathcal{B}_c\mathcal{B}D^{\ast}$, and write the Lagrangian as
\begin{align}
\begin{split}
\mathcal{L}_{\Lambda_{c}ND_{q}}&=g_{\Lambda_{c}ND_{q}}(\bar{\Lambda}_{c}i\gamma_{5}D_{q}N+h.c.),\\
\mathcal{L}_{\Lambda_{c}ND_{q}^{\ast}}&=f_{1\Lambda_{c}ND_{q}^{\ast}}(\bar{\Lambda}_{c}\gamma_{\mu}D_{q}^{\ast\mu}N+h.c.)+\frac{f_{2\Lambda_{c}ND_{q}^{\ast}}}{m_{\Lambda_{c}}+m_{N}}(\bar{\Lambda}_{c}\sigma_{\mu\nu}\partial^{\mu}D_{q}^{\ast\nu}N+h.c.)\\
\mathcal{L}_{\Sigma_{c}ND_{q}}&=g_{\Sigma_{c}ND_{q}}(\bar{\Sigma}_{c}i\gamma_{5}D_{q}N+h.c.),\\
\mathcal{L}_{\Sigma_{c}ND_{q}^{\ast}}&=f_{1\Sigma_{c}ND_{q}^{\ast}}(\bar{\Sigma}_{c}\gamma_{\mu}D_{q}^{\ast\mu}N+h.c.)+\frac{f_{2\Sigma_{c}ND_{q}^{\ast}}}{m_{\Sigma_{c}}+m_{N}}(\bar{\Sigma}_{c}\sigma_{\mu\nu}\partial^{\mu}D_{q}^{\ast\nu}N+h.c.).
\end{split}
\end{align}
where $N$ denotes baryons belong to the octet baryon matrix $\mathcal{B}$.

The strong coupling constants are taken from the literature \cite{Cheng:2004ru,Aliev:2010yx,Aliev:2010nh,Khodjamirian:2011jp,Azizi:2014bua,Yu:2016pyo,Azizi:2015tya,Ballon-Bayona:2017bwk}, and listed in Tables.  \ref{tab:VertexP}, \ref{tab:VertexV} and \ref{tab:VertexD}.
\begin{table}[h!]
\centering
\caption{\label{tab:VertexP}The strong coupling constants for $VPP$, $VVV$, $\mathcal{B}_c\mathcal{B}_cP$, $\mathcal{B}\mathcal{B}P$ and $\mathcal{B}_c\mathcal{B}D$.}

\begin{tabular}{|c|c|c|c|c|c|}
\hline
vertex & $g$ &  vertex & $g$ & vertex & $g$ \\
\hline
 $\rho\to\pi\pi$ & 6.05 & $K^{\ast}\to K\pi$ & 4.60 & $K^{\ast}\to K^{\ast}\rho$ & 5.22 \\
\hline
 $\Xi_c^0\to\Xi_c^+\pi^-$ & 0.70 & $\Xi_c^0\to\Xi_c^{\prime+}\pi^-$ & 3.10 & $\Xi_c^+\to\Xi_c^0\pi^+$ & 0.99 \\
\hline
 $\Xi_c^+\to\Xi_c^{\prime0}\pi$ & 4.38 & $\Xi_c^{\prime+}\to\Xi_c^{\prime0}\pi$ & 5.66 & $\Sigma_c^{++}\to\Xi_c^+K^+$ & $-$6.5 \\
\hline
 $\Xi_c^{\prime+}\to\Sigma_c^{++}K^-$ & 9.0 & $\Sigma_c^{++}\to\Sigma_c^+\pi^+$ & 8.0 & $\Sigma_c^{++}\to\Lambda_c^+\pi^+$ & $-$6.36\\
\hline
 $p\to n\pi^+$ & 12.0 & $\Lambda_c^+\to pD^0$ & 4.82 & $\Sigma_c^+\to pD^0$ & 3.78  \\
\hline
 $\Lambda_c^+\to nD^+$ & 4.82 & $\Sigma_c^+\to nD^+$ & 3.78 & $\Lambda_c^+\to\Xi_c^0 K^+$ & $-$0.9 \\
\hline
 $\Lambda_c^+\to\Xi_c^{\prime0}K^+$ & 4.38 & $\Lambda_c^+\to\Sigma_c^0\pi^+$ & 6.5 & $\Sigma_c^+\to\Xi_c^0K^+$ & $-$6.5 \\
\hline
 $\Sigma_c^+\to\Xi_c^{\prime0}K^+$ & 5.66 & $\Lambda^0\to\Sigma^-\pi^+$ & 10.0 & $\Xi_c^0\to\Lambda^0D^0$ & 4.82 \\
\hline
 $\Xi_c^{\prime0}\to\Lambda^0D^0$ & 3.78 & $\Xi_c^0\to\Sigma^-D^+$ & 4.82 & $\Xi_c^{\prime0}\to\Sigma^-D^+$ & 3.78 \\
\hline
\end{tabular}
\end{table}

\begin{table}[h!]
\centering
\caption{\label{tab:VertexV}The strong coupling constants for $\mathcal{B}_c\mathcal{B}_cV$, $\mathcal{B}\mathcal{B}V$ and $\mathcal{B}_c\mathcal{B}D^\ast$.}
\begin{tabular}{|c|c|c|c|c|c|}
\hline
 vertex & $f_1$ & $f_2$ & vertex & $f_1$ & $f_2$ \\
 \hline
 $\Sigma_c^{++}\to\Xi_c^+K^{\ast+}$ & $-$3.11 & $-$18.4 & $\Sigma_c^{++}\to\Xi_c^{\prime+}K^{\ast+}$ & 5.0 & 30 \\
 \hline
 $\Xi_c^+\to\Lambda_c^+\overline K^{\ast0}$ & 4.6 & 6.0 & $\Xi_c^{\prime+}\to\Lambda_c^+\overline K^{\ast0}$ & $-$2.2 & $-$13.0 \\
 \hline
 $\Sigma_c^{++}\to\Lambda_c^+\rho^+$ & $-$2.6 & $-$16.0 & $\Sigma_c^+\to\Xi_c^+\overline K^{\ast0}$ & $-$2.2 & $-$13.0 \\
 \hline
 $\Sigma_c^+\to\Xi_c^{\prime+}\overline K^{\ast0}$ & 2.82 & 19.1 & $\Sigma_c^{++}\to\Sigma_c^+\rho^+$ & 4.0 & 27.0 \\
 \hline
 $\Lambda_c^+\to pD^{\ast0}$ & 2.05 & 7.78 & $\Sigma_c^+\to pD^{\ast0}$ & 11.21 & 4.64 \\
 \hline
 $p\to n\rho^+$ & $-$2.3 & 36.1 & $\Lambda_c^+\to nD^{\ast+}$ & 2.05 & 7.78 \\
 \hline
 $\Sigma_c^+\to nD^{\ast+}$ & 11.21 & 4.64 & $\Xi_c^0\to\Lambda_c^+K^{\ast-}$ & $-$4.6 & $-$6.0 \\
 \hline
 $\Xi_c^{\prime0}\to\Lambda_c^+K^{\ast-}$ & 2.12 & 15.6 & $\Sigma_c^0\to\Xi_c^0K^{\ast0}$ & $-$3.11 & $-$18.4 \\
 \hline
 $\Sigma_c^0\to\Xi_c^{\prime0}K^{\ast0}$ & 5.0 & 30.0 & $\Sigma_c^+\to\Lambda_c^+\rho^-$ & 2.6 & 16.0 \\
 \hline
 $\Xi_c^0\to\Xi_c^0\rho^0$ & $-$6.0 & $-$7.5 & $\Xi_c^{\prime0}\to\Xi_c^0\rho^0$ & 2.12 & 15.6 \\
 \hline
 $\Xi_c^{\prime0}\to\Xi_c^{\prime0}\rho^0$ & $-$2.5 & $-$16.0 & $\Xi_c^+\to\Xi_c^0\rho^+$ & 8.5 & 10.6 \\
 \hline
 $\Xi_c^{\prime0}\to\Xi_c^{\prime+}\rho^-$ & 2.12 & 15.6 & $\Lambda^0\to\Sigma^-\rho^+$ & 2.0 & 12.3 \\
 \hline
 $\Xi_c^0\to\Lambda^0D^{\ast0}$ & 2.05 & 7.78 & $\Xi_c^{\prime0}\to\Lambda^0D^{\ast0}$ & 11.21 & 4.64  \\
 \hline
\end{tabular}
\end{table}

\begin{table}[h!]
\centering
\caption{\label{tab:VertexD}The strong coupling constants for $D^{\ast}D\Pi$, $DDV$ and $D^{\ast}D^{\ast}V$.}

\begin{tabular}{|c|c|c|c|c|c|c|}
\hline
vertex & $g$ &  vertex & $g$ & vertex & $g$ &$f$\\
\hline
 $D^\ast\to D\pi$ & 17.9 & $D\to D\rho$ & 3.69 & $D^\ast\to D^\ast\rho$ & 3.69 & 4.61 \\
\hline
\end{tabular}
\end{table}


\begin{thebibliography}{99}

\bibitem{Klempt:2009pi}
  E.~Klempt and J.~M.~Richard,
  Rev.\ Mod.\ Phys.\  {\bf 82}, 1095 (2010)
  doi:10.1103/RevModPhys.82.1095
  [arXiv:0901.2055 [hep-ph]].


\bibitem{Crede:2013sze}
  V.~Crede and W.~Roberts,
  Rept.\ Prog.\ Phys.\  {\bf 76}, 076301 (2013)
  doi:10.1088/0034-4885/76/7/076301
  [arXiv:1302.7299 [nucl-ex]].


\bibitem{Cheng:2015iom}
  H.~Y.~Cheng,
  Front.\ Phys.\ (Beijing) {\bf 10}, no. 6, 101406 (2015).
  doi:10.1007/s11467-015-0483-z


\bibitem{Chen:2016spr}
  H.~X.~Chen, W.~Chen, X.~Liu, Y.~R.~Liu and S.~L.~Zhu,
  Rept.\ Prog.\ Phys.\  {\bf 80}, no. 7, 076201 (2017)
  doi:10.1088/1361-6633/aa6420
  [arXiv:1609.08928 [hep-ph]].


\bibitem{Mattson:2002vu}
  M.~Mattson {\it et al.} [SELEX Collaboration],
  Phys.\ Rev.\ Lett.\  {\bf 89}, 112001 (2002)
  doi:10.1103/PhysRevLett.89.112001
  [hep-ex/0208014].


\bibitem{Ocherashvili:2004hi}
  A.~Ocherashvili {\it et al.} [SELEX Collaboration],
  Phys.\ Lett.\ B {\bf 628}, 18 (2005)
  doi:10.1016/j.physletb.2005.09.043
  [hep-ex/0406033].


\bibitem{Ratti:2003ez}
  S.~P.~Ratti,
  Nucl.\ Phys.\ Proc.\ Suppl.\  {\bf 115}, 33 (2003).
  doi:10.1016/S0920-5632(02)01948-5


\bibitem{Aubert:2006qw}
  B.~Aubert {\it et al.} [BaBar Collaboration],
  Phys.\ Rev.\ D {\bf 74}, 011103 (2006)
  doi:10.1103/PhysRevD.74.011103
  [hep-ex/0605075].


\bibitem{Chistov:2006zj}
  R.~Chistov {\it et al.} [Belle Collaboration],
  Phys.\ Rev.\ Lett.\  {\bf 97}, 162001 (2006)
  doi:10.1103/PhysRevLett.97.162001
  [hep-ex/0606051].


\bibitem{Kato:2013ynr}
  Y.~Kato {\it et al.} [Belle Collaboration],
  Phys.\ Rev.\ D {\bf 89}, no. 5, 052003 (2014)
  doi:10.1103/PhysRevD.89.052003
  [arXiv:1312.1026 [hep-ex]].


\bibitem{Aaij:2013voa}
  R.~Aaij {\it et al.} [LHCb Collaboration],
  JHEP {\bf 1312}, 090 (2013)
  doi:10.1007/JHEP12(2013)090
  [arXiv:1310.2538 [hep-ex]].


\bibitem{Lewis:2001iz}
  R.~Lewis, N.~Mathur and R.~M.~Woloshyn,
  Phys.\ Rev.\ D {\bf 64}, 094509 (2001)
  doi:10.1103/PhysRevD.64.094509
  [hep-ph/0107037].


\bibitem{Flynn:2003vz}
  J.~M.~Flynn {\it et al.} [UKQCD Collaboration],
  JHEP {\bf 0307}, 066 (2003)
  doi:10.1088/1126-6708/2003/07/066
  [hep-lat/0307025].


\bibitem{Liu:2009jc}
  L.~Liu, H.~W.~Lin, K.~Orginos and A.~Walker-Loud,
  Phys.\ Rev.\ D {\bf 81}, 094505 (2010)
  doi:10.1103/PhysRevD.81.094505
  [arXiv:0909.3294 [hep-lat]].


\bibitem{Alexandrou:2012xk}
  C.~Alexandrou, J.~Carbonell, D.~Christaras, V.~Drach, M.~Gravina and M.~Papinutto,
  Phys.\ Rev.\ D {\bf 86}, 114501 (2012)
  doi:10.1103/PhysRevD.86.114501
  [arXiv:1205.6856 [hep-lat]].


\bibitem{Briceno:2012wt}
  R.~A.~Briceno, H.~W.~Lin and D.~R.~Bolton,
  Phys.\ Rev.\ D {\bf 86}, 094504 (2012)
  doi:10.1103/PhysRevD.86.094504
  [arXiv:1207.3536 [hep-lat]].


\bibitem{Alexandrou:2014sha}
  C.~Alexandrou, V.~Drach, K.~Jansen, C.~Kallidonis and G.~Koutsou,
  Phys.\ Rev.\ D {\bf 90}, no. 7, 074501 (2014)
  doi:10.1103/PhysRevD.90.074501
  [arXiv:1406.4310 [hep-lat]].


\bibitem{Zhang:2011hi}
  J.~W.~Zhang, X.~G.~Wu, T.~Zhong, Y.~Yu and Z.~Y.~Fang,
  Phys.\ Rev.\ D {\bf 83}, 034026 (2011)
  doi:10.1103/PhysRevD.83.034026
  [arXiv:1101.1130 [hep-ph]].


\bibitem{Chang:2005bf}
  C.~H.~Chang, C.~F.~Qiao, J.~X.~Wang and X.~G.~Wu,
  Phys.\ Rev.\ D {\bf 71}, 074012 (2005)
  doi:10.1103/PhysRevD.71.074012
  [hep-ph/0502155].


\bibitem{Karliner:2014gca}
  M.~Karliner and J.~L.~Rosner,
  Phys.\ Rev.\ D {\bf 90}, no. 9, 094007 (2014)
  doi:10.1103/PhysRevD.90.094007
  [arXiv:1408.5877 [hep-ph]].


\bibitem{Kiselev:2001fw}
  V.~V.~Kiselev and A.~K.~Likhoded,
  Phys.\ Usp.\  {\bf 45}, 455 (2002)
  [Usp.\ Fiz.\ Nauk {\bf 172}, 497 (2002)]
  doi:10.1070/PU2002v045n05ABEH000958
  [hep-ph/0103169].


\bibitem{Chang:2007xa}
  C.~H.~Chang, T.~Li, X.~Q.~Li and Y.~M.~Wang,
  Commun.\ Theor.\ Phys.\  {\bf 49}, 993 (2008)
  doi:10.1088/0253-6102/49/4/38
  [arXiv:0704.0016 [hep-ph]].


\bibitem{Onishchenko:2000yp}
  A.~I.~Onishchenko,
  hep-ph/0006295.


\bibitem{Guberina:1999mx}
  B.~Guberina, B.~Melic and H.~Stefancic,
  Eur.\ Phys.\ J.\ C {\bf 9}, 213 (1999)
  [Eur.\ Phys.\ J.\ C {\bf 13}, 551 (2000)]
  doi:10.1007/s100529900039, 10.1007/s100520050525
  [hep-ph/9901323].


\bibitem{Ke:2007tg}
  H.~W.~Ke, X.~Q.~Li and Z.~T.~Wei,
  Phys.\ Rev.\ D {\bf 77}, 014020 (2008)
  doi:10.1103/PhysRevD.77.014020
  [arXiv:0710.1927 [hep-ph]].


\bibitem{Ke:2012wa}
  H.~W.~Ke, X.~H.~Yuan, X.~Q.~Li, Z.~T.~Wei and Y.~X.~Zhang,
  Phys.\ Rev.\ D {\bf 86}, 114005 (2012)
  doi:10.1103/PhysRevD.86.114005
  [arXiv:1207.3477 [hep-ph]].


\bibitem{Cheng:1996if}
  H.~Y.~Cheng, C.~Y.~Cheung and C.~W.~Hwang,
  Phys.\ Rev.\ D {\bf 55}, 1559 (1997)
  doi:10.1103/PhysRevD.55.1559
  [hep-ph/9607332].

\bibitem{Li:2017ndo}
  R.~H.~Li, C.~D.~L眉, W.~Wang, F.~S.~Yu and Z.~T.~Zou,
  Phys.\ Lett.\ B {\bf 767}, 232 (2017)
  doi:10.1016/j.physletb.2017.02.003
  [arXiv:1701.03284 [hep-ph]].

\bibitem{1707.02834}
  W.~Wang, F.~S.~Yu and Z.~X.~Zhao,
  Eur.\ Phys.\ J.\ C {\bf 77}, no. 11, 781 (2017)
  doi:10.1140/epjc/s10052-017-5360-1
  [arXiv:1707.02834 [hep-ph]].

\bibitem{Wang:2017azm}
  W.~Wang, Z.~P.~Xing and J.~Xu,
  Eur.\ Phys.\ J.\ C {\bf 77}, no. 11, 800 (2017)
  doi:10.1140/epjc/s10052-017-5363-y
  [arXiv:1707.06570 [hep-ph]].

\bibitem{Shi:2017dto}
  Y.~J.~Shi, W.~Wang, Y.~Xing and J.~Xu,
  Eur.\ Phys.\ J.\ C {\bf 78}, no. 1, 56 (2018)
  doi:10.1140/epjc/s10052-018-5532-7
  [arXiv:1712.03830 [hep-ph]].

\bibitem{Li:2012cfa}
  H.~n.~Li, C.~D.~Lu and F.~S.~Yu,
  Phys.\ Rev.\ D {\bf 86}, 036012 (2012)
  doi:10.1103/PhysRevD.86.036012
  [arXiv:1203.3120 [hep-ph]].


\bibitem{Link:2001rn}
  J.~M.~Link {\it et al.} [FOCUS Collaboration],
  Phys.\ Lett.\ B {\bf 512}, 277 (2001)
  doi:10.1016/S0370-2693(01)00590-1
  [hep-ex/0102040].


\bibitem{Link:2002zx}
  J.~M.~Link {\it et al.} [FOCUS Collaboration],
  Phys.\ Lett.\ B {\bf 540}, 25 (2002)
  doi:10.1016/S0370-2693(02)02103-2
  [hep-ex/0206013].


\bibitem{Ablikim:2002ep}
  M.~Ablikim, D.~S.~Du and M.~Z.~Yang,
  Phys.\ Lett.\ B {\bf 536}, 34 (2002)
  doi:10.1016/S0370-2693(02)01812-9
  [hep-ph/0201168].


\bibitem{Li:2002pj}
  J.~W.~Li, M.~Z.~Yang and D.~S.~Du,
  HEPNP {\bf 27}, 665 (2003)
  [hep-ph/0206154].


\bibitem{Fajfer:2003ag}
  S.~Fajfer, A.~Prapotnik, P.~Singer and J.~Zupan,
  Phys.\ Rev.\ D {\bf 68}, 094012 (2003)
  doi:10.1103/PhysRevD.68.094012
  [hep-ph/0308100].


\bibitem{Li:1997vu}
  X.~Q.~Li and B.~S.~Zou,
  Phys.\ Rev.\ D {\bf 57}, 1518 (1998)
  doi:10.1103/PhysRevD.57.1518
  [hep-ph/9709508].


\bibitem{Chen:2002jr}
  S.~L.~Chen, X.~H.~Guo, X.~Q.~Li and G.~L.~Wang,
  Commun.\ Theor.\ Phys.\  {\bf 40}, 563 (2003)
  doi:10.1088/0253-6102/40/5/563
  [hep-ph/0208006].


\bibitem{Ablikim:2017ors}
  M.~Ablikim {\it et al.} [BESIII Collaboration],
  Phys.\ Rev.\ D {\bf 95}, no. 11, 111102 (2017)
  doi:10.1103/PhysRevD.95.111102
  [arXiv:1702.05279 [hep-ex]].


\bibitem{Cheng:2004ru}
  H.~Y.~Cheng, C.~K.~Chua and A.~Soni,
  Phys.\ Rev.\ D {\bf 71}, 014030 (2005)
  doi:10.1103/PhysRevD.71.014030
  [hep-ph/0409317].


\bibitem{Yan:1992gz}
  T.~M.~Yan, H.~Y.~Cheng, C.~Y.~Cheung, G.~L.~Lin, Y.~C.~Lin and H.~L.~Yu,
  Phys.\ Rev.\ D {\bf 46}, 1148 (1992)
  Erratum: [Phys.\ Rev.\ D {\bf 55}, 5851 (1997)].
  doi:10.1103/PhysRevD.46.1148, 10.1103/PhysRevD.55.5851


\bibitem{Casalbuoni:1996pg}
  R.~Casalbuoni, A.~Deandrea, N.~Di Bartolomeo, R.~Gatto, F.~Feruglio and G.~Nardulli,
  Phys.\ Rept.\  {\bf 281}, 145 (1997)
  doi:10.1016/S0370-1573(96)00027-0
  [hep-ph/9605342].


\bibitem{Meissner:1987ge}
  U.~G.~Meissner,
  Phys.\ Rept.\  {\bf 161}, 213 (1988).
  doi:10.1016/0370-1573(88)90090-7


\bibitem{Li:2012bt}
  N.~Li and S.~L.~Zhu,
  Phys.\ Rev.\ D {\bf 86}, 014020 (2012)
  doi:10.1103/PhysRevD.86.014020
  [arXiv:1204.3364 [hep-ph]].


\bibitem{Aliev:2010yx}
  T.~M.~Aliev, K.~Azizi and M.~Savci,
  Phys.\ Lett.\ B {\bf 696}, 220 (2011)
  doi:10.1016/j.physletb.2010.12.027
  [arXiv:1009.3658 [hep-ph]].


\bibitem{Aliev:2010nh}
  T.~M.~Aliev, K.~Azizi and M.~Savci,
  Nucl.\ Phys.\ A {\bf 852}, 141 (2011)
  doi:10.1016/j.nuclphysa.2011.01.011
  [arXiv:1011.0086 [hep-ph]].


\bibitem{Khodjamirian:2011jp}
  A.~Khodjamirian, C.~Klein, T.~Mannel and Y.-M.~Wang,
  JHEP {\bf 1109}, 106 (2011)
  doi:10.1007/JHEP09(2011)106
  [arXiv:1108.2971 [hep-ph]].


\bibitem{Azizi:2014bua}
  K.~Azizi, Y.~Sarac and H.~Sundu,
  Phys.\ Rev.\ D {\bf 90}, no. 11, 114011 (2014)
  doi:10.1103/PhysRevD.90.114011
  [arXiv:1410.7548 [hep-ph]].


\bibitem{Yu:2016pyo}
  G.~L.~Yu, Z.~G.~Wang and Z.~Y.~Li,
  Chin.\ Phys.\ C {\bf 41}, no. 8, 083104 (2017)
  doi:10.1088/1674-1137/41/8/083104
  [arXiv:1608.03460 [hep-ph]].


\bibitem{Azizi:2015tya}
  K.~Azizi, Y.~Sarac and H.~Sundu,
  Nucl.\ Phys.\ A {\bf 943}, 159 (2015)
  doi:10.1016/j.nuclphysa.2015.09.005
  [arXiv:1501.05084 [hep-ph]].


\bibitem{Ballon-Bayona:2017bwk}
  A.~Ballon-Bayona, G.~Krein and C.~Miller,
  Phys.\ Rev.\ D {\bf 96}, no. 1, 014017 (2017)
  doi:10.1103/PhysRevD.96.014017
  [arXiv:1702.08417 [hep-ph]].


\bibitem{Cheng:2010ry}
  H.~Y.~Cheng and C.~W.~Chiang,
  Phys.\ Rev.\ D {\bf 81}, 074021 (2010)
  doi:10.1103/PhysRevD.81.074021
  [arXiv:1001.0987 [hep-ph]].


\bibitem{Cheng:2010rv}
  H.~Y.~Cheng and C.~W.~Chiang,
  Phys.\ Rev.\ D {\bf 81}, 114020 (2010)
  doi:10.1103/PhysRevD.81.114020
  [arXiv:1005.1106 [hep-ph]].


\bibitem{Aaij:2017ueg}
  R.~Aaij {\it et al.} [LHCb Collaboration],
  Phys.\ Rev.\ Lett.\  {\bf 119}, no. 11, 112001 (2017)
  doi:10.1103/PhysRevLett.119.112001
  [arXiv:1707.01621 [hep-ex]].

    \end{thebibliography}
\end{document}